\documentstyle[twocolumn,aps,epsfig]{revtex}

\begin{document}
\newcommand{\mb}[1]{\mbox{\boldmath $#1$}}
\draft

\title{
Mott Transition
in the Two-Dimensional Flux Phase
}
\author{
Y.Otsuka$^1$
and Y.Hatsugai$^{1,2}$
}
\address{
$^1$Department of Applied Physics, University of Tokyo,
7-3-1 Hongo Bunkyo-ku,
Tokyo 113-8656, Japan\\
$^2$PRESTO,
Japan Science and Technology Corporation,
Kawaguchi-shi,Saitama, 332-0012, Japan
}

\maketitle
\begin{abstract}
{
  Effects of the electron-electron interaction
in the two-dimensional flux phase are investigated.
  We treat the half-filled Hubbard model 
with a magnetic flux $\pi$ per plaquette
by the quantum Monte Carlo method.
  When the interaction is small,
an antiferromagnetic long-range does not exist
and the charge fluctuation is different from 
that of the Mott insulator
  It suggests that the Mott transition occurs 
at finite strength of the interaction in the flux phase, 
which is in contrast to the standard Hubbard model.
}
\end{abstract}
\section{introduction}
  There has been a proposal that
an order takes place on a link 
in several interacting lattice-fermion systems.
 Especially when the link order has a phase factor,
it brings an effective magnetic field.
  Sometime the order can be topological 
in the sense that
the phase factor itself is not a well defined order parameter 
but the flux characterizes the phase.
  One of such systems is the flux phase
which was proposed to describe
the ground state properties of 
several interacting lattice-fermion systems
e.g.
the Hubbard model, 
the t-J model, 
and their variants
\cite{ref-am1,ref-azha1,ref-kl1,ref-dhl1,ref-ilw1,ref-pwl1,ref-mws1}.
  Recently there has been a resurgence of interest
in the flux phase
and evidences are accumulating for its reality.
  For example, 
it was revealed a bond order takes place 
in the one-dimensional extended Hubbard model 
at half filling 
\cite{ref-mn1} 
which can be understood by 
the one-dimensional analog of the flux phase.
  Further, it has been discussed that
a hidden topological order exists in cuprates,
which shares some aspects with the flux phase 
\cite{ref-srdc1}.
  In general, however,
the flux phase competes with other instabilities 
e.g.
superconductivity, 
antiferromagnetism, 
charge order
and localization due to disorder
(see refs. 
\cite{ref-ff1,ref-hl1,ref-hwk1,ref-yy1} 
for the effects of disorder on the flux phase).
  In particular,
the flux phase often competes with superconductivity,
which is a direct consequence from the SU(2) symmetry 
at half filling
\cite{ref-azha1}.
  We also note that
the flux phase can emerge dynamically 
\cite{ref-ilw1,ref-pwl1}
as well as in a static form 
\cite{ref-srdc1}.

  In this paper, we report
effects of the interaction 
in the two-dimensional flux phase.
  We choose the two-dimensional Hubbard model
with a magnetic flux $\phi=\pi$ per square plaquette 
and compare the results 
with the standard Hubbard model ($\phi=0$).
  In the standard Hubbard model at half filling,
it is believed that 
an infinitesimally small interaction drives 
the ground state to the Mott insulator,
where a finite charge gap opens 
and an antiferromagnetic long-range order exists.
  This is consistent with the nesting argument 
in the weak coupling region.
  On the the hand, in the flux phase without the interaction, 
the density of state disappear linearly at the Fermi energy,
which suggests that
the structure of the low energy excitations is singular 
as compared with a simple Fermi liquid 
and the nesting instability is absent.
  Therefore one can expect 
an interaction-driven quantum phase transition
from a singular quantum liquid
(density of state is linearly vanishing without interaction)
to a gapped insulator (Mott insulator).

\section{flux phase}

  The flux phase is given by the ground state
of the following simple Hamiltonian,
\begin{eqnarray}
   \label{eq:flux-model}
   {\cal H}_{\rm F} &=&
   \sum _{ \langle j,k \rangle , \sigma}
   (
   c_{j \sigma }^{\dagger}
   t_{jk}
   c_{k \sigma} +
   c_{k \sigma}^{\dagger}
   t_{kj}^{*}
   c_{j \sigma}
   ),
\end{eqnarray}
where
$\langle j,k \rangle$ denotes a nearest-neighbor link. 
  The amplitude of $t_{jk}$ is constant 
but its phase factor 
$t_{jk}/|t_{jk}|=e^{i\theta_{jk}}$
satisfies a condition 
$ \phi = \sum_{{\rm plaquette}} \theta_{jk}$.
  It leads to a uniform magnetic flux per plaquette. 
  The phase factor $\theta _{jk}$ itself is not fixed 
but the flux $\phi$ is fixed,
which is the gauge independent quantity.
  This Hamiltonian was proposed 
as an  effective model (in the mean field level) 
of several correlated electron systems
and discussed in many different contexts
\cite{ref-am1,ref-azha1,ref-kl1,ref-dhl1,ref-ilw1,ref-pwl1,ref-mws1,ref-hl1,ref-hwk1,ref-yy1,ref-hlrw,ref-hhkm,ref-ehl1}.
  One of the focuses was the stability of
the flux state
\cite{ref-hlrw,ref-hhkm,ref-ehl1}. 
  Following the discussion, 
the optimum, energy-minimizing, magnetic flux 
at half filling (this is the simplest case)
is considered as $\pi$ per square plaquette
\cite{ref-hlrw,ref-hhkm}.
  Furthermore, 
we note that Lieb gave some rigorous results
for the stability in a general form \cite{ref-ehl1}.

  At half filling,
the low-lying excitations
of the flux phase
is described by massless Dirac fermions.
  There is a gauge freedom 
for the phase factor $\theta_{jk}$
but let us fix them by choosing as
$t_{j+{\hat x},j}=(-1)^{j_{y}}t$, $t_{j+{\hat y},j}=t$
and otherwise zero
where
$j=(j_{x},j_{y}){\in}{\bf Z}^{2}$,
${\hat x}=(1,0)$,
${\hat y}=(0,1)$.
  The energy bands in this gauge are given by
\begin{eqnarray}
E({\bf k})
&=&
\pm 2t\sqrt {{\cos}^{2}k_{x}+{\cos}^{2}k_{y}}\\
&\approx&
\pm 2t\sqrt {(k_{x}-k_{x}^{i})^{2}+(k_{y}-k_{y}^{i})^{2}}\ (i=1,2),
\end{eqnarray}
where
$(k_{x},k_{y}){\in}[-{\pi},{\pi}){\times}[0,{\pi})$,
${\bf k}^{1} = (k_{x}^{1},k_{y}^{1})=({\pi}/2,{\pi}/2)$
and
${\bf k}^{2} = (k_{x}^{2},k_{y}^{2})=(-{\pi}/2,{\pi}/2)$.
  Therefore the low-lying excitations
are described by massless Dirac fermions
at these two gap-closing points
and the density of states $D(\epsilon)$ near the Fermi energy
vanishes linearly, $D(\epsilon) \propto |\epsilon|$.
  The density of states $D(\epsilon)$ is singular and
it leads to the suppression
of the instability against the Mott insulator
as discussed below.
  Note that the dispersion is gauge dependent but the
density of state is gauge independent.
  We focus on only the gauge independent quantity 
in this paper.

\section{model and method}

  We investigate effects of the interaction 
in the flux phase
by the following Hamiltonian,
\begin{eqnarray}
  \label{eq:model_Hamiltonian}
   {\cal H} &=&
   \sum _{ \langle j,k \rangle , \sigma}
   (
   c_{j \sigma }^{\dagger}
   t_{jk}
   c_{k \sigma} +
   c_{k \sigma}^{\dagger}
   t_{kj}^{*}
   c_{j \sigma}
   )
   \nonumber
   \\
  &+& U \sum_{i}
   (n_{i \uparrow}-1/2)(n_{i \downarrow}-1/2),
\end{eqnarray}
where 
$\langle j,k \rangle$ denotes a nearest-neighbor link
and $U$ a on-site Coulomb repulsion.  
  The geometry is set to be a two-dimensional square lattice 
and a periodic boundary condition is imposed.
  The grand-canonical ensemble is employed 
and we put the system half-filled 
by the particle-hole symmetry.
  The $|t_{jk}|$ is set to be constant ($=t$) and,
based on the Lieb's theorem\cite{ref-ehl1},
the phase factor $e^{i \theta_{jk}}$ is chosen so that 
the magnetic flux $\phi$ is $\pi$ per plaquette
i.e. $\pi$-flux phase 
($\phi \equiv \sum_{{\rm plaquette}} \theta_{jk}= \pi$).
  We always tries to compare 
the results of the flux phase ($\phi=\pi$)
with those of the standard Hubbard model ($\phi=0$).
  It is to be noted that 
neither the translational nor time-reversal symmetry 
is broken in the $\pi$-flux phase.

  In order to study the system 
based on a non-perturbative approximation free method,
the quantum Monte Carlo (QMC) technique is applied 
\cite{ref-qmc1,ref-qmc2}.
  We use the grand-canonical scheme at finite temperatures.
  Due to the particle-hole symmetry in the Hamiltonian
(\ref{eq:model_Hamiltonian}), 
the negative-sign problem does not occur.
  The simulations were performed on a square lattice 
with a size up to $N = 12 \times 12$ 
at a temperature down to $T=0.05t$.  
  The Trotter decomposition is performed 
in the imaginary-time direction 
and the time slice is $\Delta\tau \simeq 0.10/t$.
  We have checked that 
the systematic errors due to the Trotter decomposition
are almost independent of temperatures
and does not change the essential features after the extrapolation.
  We have typically performed 500 Monte Carlo sweeps
in order to reach a thermal equilibrium 
followed by 5000 measurement sweeps.
  The measurements are divided into 10 blocks 
and the statistical error is defined 
by the variance among the blocks.

  The Mott insulator is characterized 
by the following two features.
  One is a strong suppression of the charge fluctuation 
and the other is a presence of 
the strong antiferromagnetic spin correlation.
  In order to detect signals of the Mott transition,
we have calculated 
the charge compressibility
and the magnetic structure factor.
The charge compressibility is defined by
\begin{eqnarray}
   \label{eq:compressibility}
    \kappa =
    \frac{1}{N}
    \frac{\partial N_{e}}{\partial \mu}
    =
    \frac{\beta}{N}
    \left(
	\langle N_{e}^{2} \rangle
	-
	\langle N_{e} \rangle ^{2}
   \right),
\end{eqnarray}
where
$N_{e}$ is the number of electrons
and $\beta$ an inverse temperature.
  The charge compressibility $\kappa$ measures 
the charge fluctuation directly. 
  If the system has a finite charge gap, 
$\kappa$ shows a thermally-activated behavior 
and vanishes at zero temperature.
  The magnetic structure factor is given by
\begin{eqnarray}
   \label{eq:magnetic_structure_factor}
    && S(\mb{q}) \nonumber \\ && =
   \frac{1}{N}
   \sum_{i,j}
   e^{ i\mb{q} \cdot ( \mb{r}_{i}- \mb{r}_{j})}
   \langle
   ( n_{i \uparrow} - n_{i \downarrow} )
   ( n_{j \uparrow} - n_{j \downarrow} )
   \rangle.
\end{eqnarray}
  If the system has an antiferromagnetic long-range order,
$S(\pi,\pi)$ shows a diverging behaver 
as the temperature decreases.

\section{results}
\begin{figure}[htbp]
 \begin{center}
  \leavevmode
  \epsfig{file=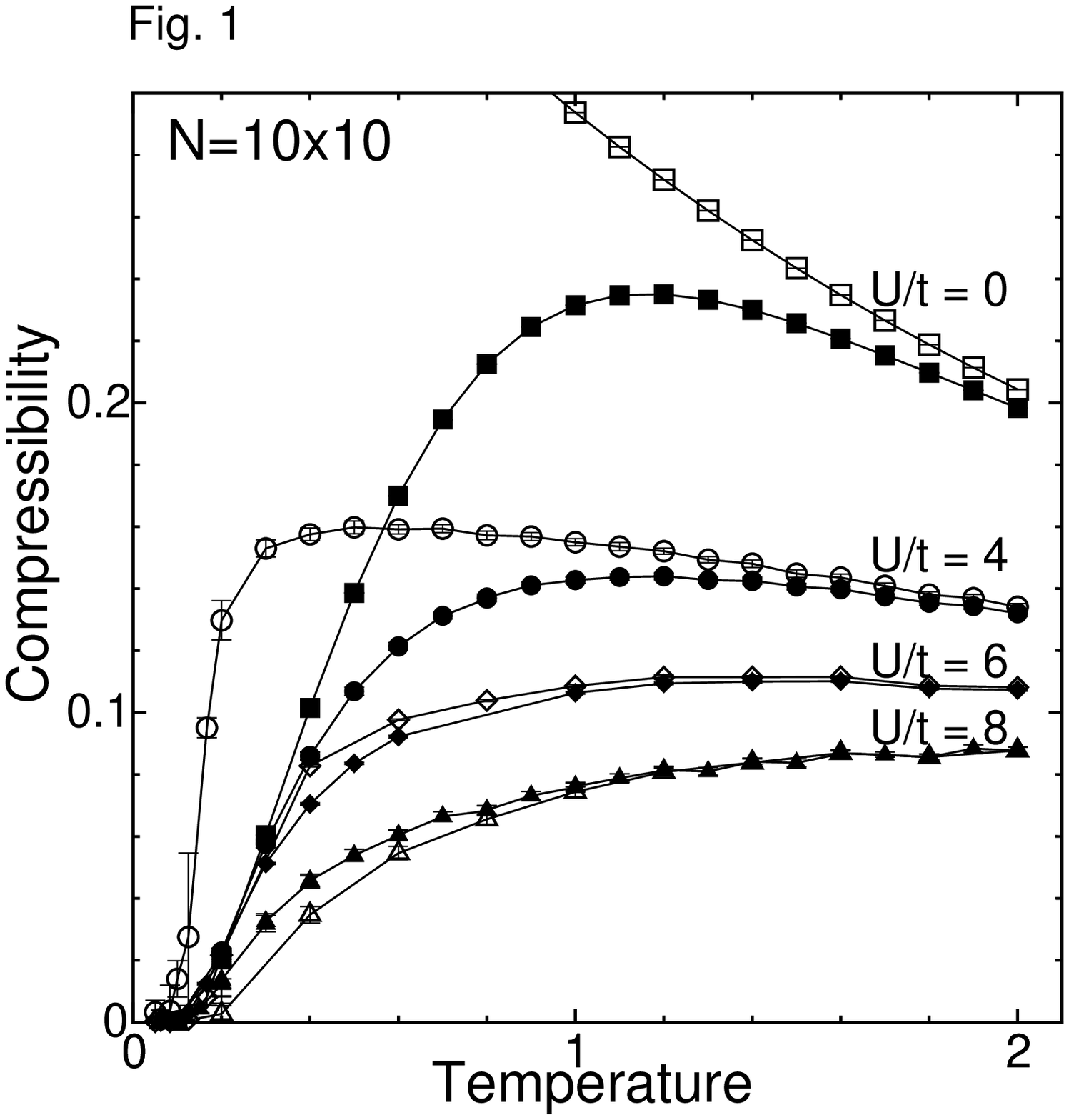,width=6.0cm}
  \caption{
  Temperature dependence of the compressibility
  for a $N=10 \times 10$ lattice
  with interaction strengths
  $U/t$=
  0 (squares),
  4 (circles),
  6 (diamonds), and
  8 (triangles).
  Solid symbols are for $\phi=\pi$ 
  and open symbols for $\phi=0$.
  When the interaction is small ($U/t \le 4$),
  the difference between $\phi =0$ and $\phi=\pi$
  is large over wide range of temperatures.
  }
  \label{fig:charge-01}
 \end{center}
\end{figure}

  First let us discuss effects of the interaction
on the charge compressibility.
  We compare the result with those of the standard Hubbard model 
to clarify the effects of the flux,
that is, 
the structure of the low energy excitations.
  Fig.\ref{fig:charge-01} shows results 
of the charge compressibility $\kappa$.  
  Since the $\pi$-flux state at $U/t=0$ has 
gapless points in the Brillouin zone 
and the density of states $D(\epsilon)$ near the Fermi energy
vanishes linearly, $D(\epsilon) \propto |\epsilon|$, 
the compressibility $\kappa$ does {\it not} show
thermally-activated behavior. 
  Even at $U/t=4$, our data show that
effects of the flux is still relevant 
and the charge gap is not well defined
in the flux phase.
  This implies that 
the singular spectrum of the excitations seems to 
survive even with the interaction.
  Then this possible phase is clearly 
not a simple Fermi liquid 
but a singular quantum liquid.
  As the strength of the interaction increases,
effects of the flux becomes irrelevant.
  If the interaction is sufficiently strong ($U \gg t$),
the system becomes the Mott insulator 
with a finite charge gap 
which is the order of the interaction.
  Therefore the results suggest
the existence of a finite value of the interaction strength, $U_c$,
which separates from a gapless singular phase from the gapped one.

\begin{figure}[htbp]
 \begin{center}
  \leavevmode
  \epsfig{file=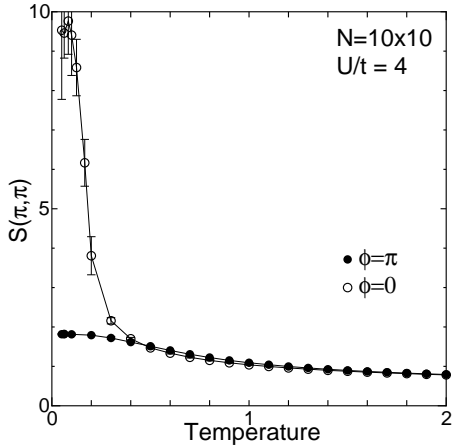,width=6.0cm}
  \caption{
  The antiferromagnetic structure factor $S(\pi, \pi)$
  as a function of temperature for $U/t=4$
  on a $N=10 \times 10$ lattice.
  In the case of $\phi=0$,
  $S(\pi, \pi)$ diverges at low temperatures
  due to the formation of the antiferromagnetic order.
  On the other hand,  for $\phi=\pi$,
  $S(\pi, \pi)$ does not show a diverging behavior.}
  \label{fig:spin-01}
 \end{center}
\end{figure}

\begin{figure}[htbp]
 \begin{center}
  \leavevmode
  \epsfig{file=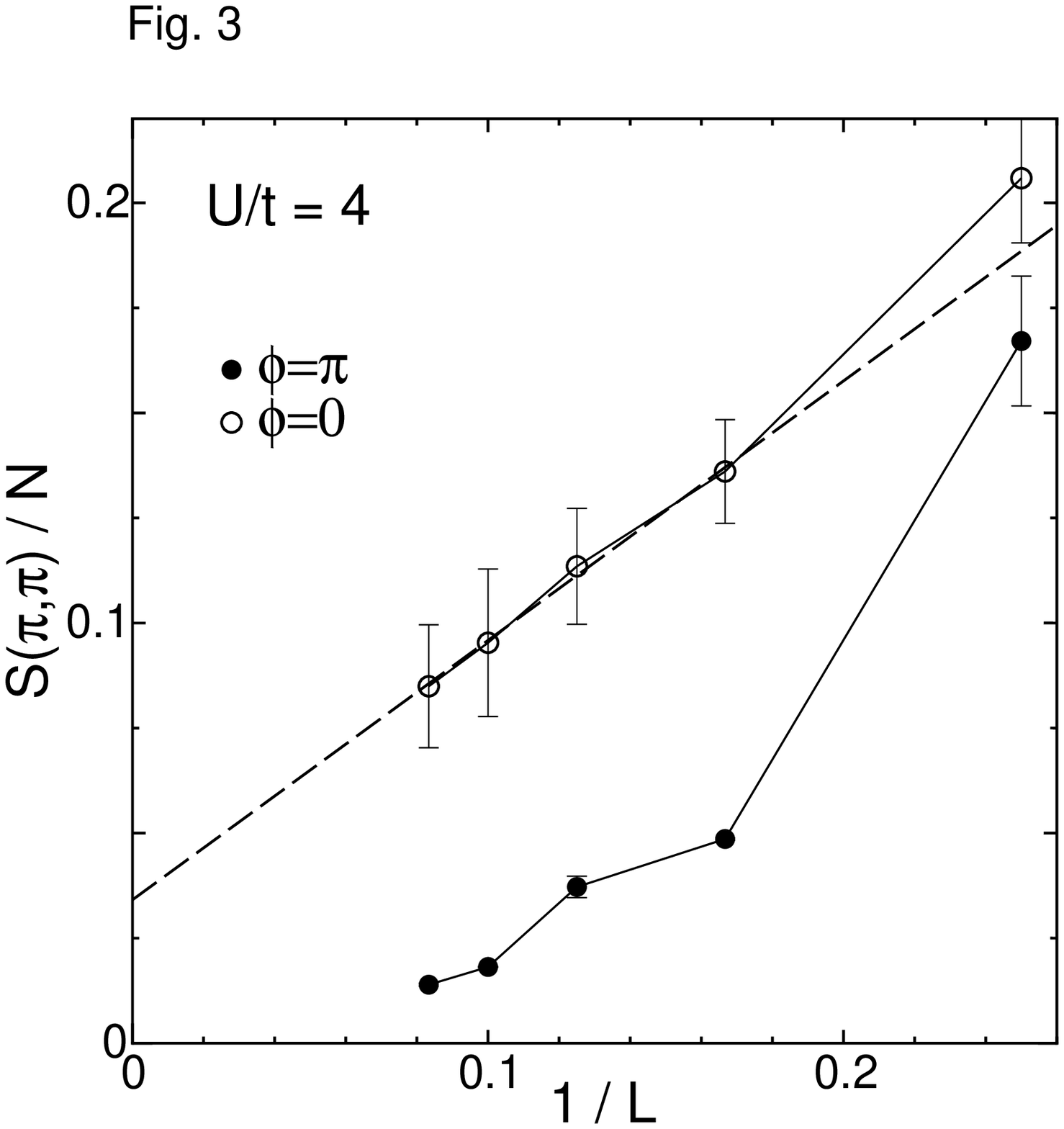,width=6.0cm}
  \caption{
  Extrapolation of antiferromagnetic long-range order.
  The dashed line is a least-squares fit 
  to the data for $\phi=0$.
  For $\phi=0$, the points extrapolate to a finite value,
  indicating that the ground state has 
  an antiferromagnetic long-range order.
  On the other hand, for $\phi=\pi$, 
  $S(\pi, \pi)$ versus $1/L$ 
  suggests the absent of an antiferromagnetic long-range order.
  }
  \label{fig:spin-02}
 \end{center}
\end{figure}

  Fig.\ref{fig:spin-01} shows 
the antiferromagnetic structure factor $S(\pi,\pi)$ 
versus temperatures.  
  For the standard half-filled Hubbard model,
since the ground state has an antiferromagnetic long-range order,
$S(\pi,\pi)$ shows a diverging behavior as the temperature decreases
(it saturates when the antiferromagnetic correlation length is longer then
the lattice size).
  On the other hand, for the flux phase, 
the formation of the long-range antiferromagnetic order 
is not observed for $U/t \le 4$.
  According to the spin-wave theory, $S(\pi,\pi)$
at the zero temperature increases with a lattice size as
\begin{equation}
   \label{eq:extrapolation} \frac{S(\pi,\pi)}{N} = \frac{m^{2}}{3} +
   O(N^{-1/2}),
\end{equation}
with $m$ the staggered magnetization 
which is an order parameter of an antiferromagnetic long-range order.  
  Using this relation, 
we try to obtain $m^{2}$ 
by plotting $S(\pi,\pi)/N$ versus $N^{-1/2}=L^{-1}$.
  Fig.\ref{fig:spin-02} shows the plots for $U/t=4$.
  The temperature is set to be $T=0.05t$ 
where the system reaches
the zero temperature limit for the system size.
  For the standard Hubbard model, 
the data follow the relation
(\ref{eq:extrapolation}) 
and the extrapolation value is finite
indicating the existence of the antiferromagnetic long-range order.
  On the other hand, for the flux phase, 
the relation
(\ref{eq:extrapolation}) 
with $m>0$ 
does not hold i.e. 
there is no antiferromagnetic long-range order, 
which is in contrast to the standard Hubbard model.
When one discuss perturbatively,
in the flux phase,
the stoner instability is strongly suppressed
due to the absence of the low energy excitations.
The numerical results are consistent with 
this discussion at least in the weak coupling.

\begin{figure}[htbp]
 \begin{center}
  \leavevmode
  \epsfig{file=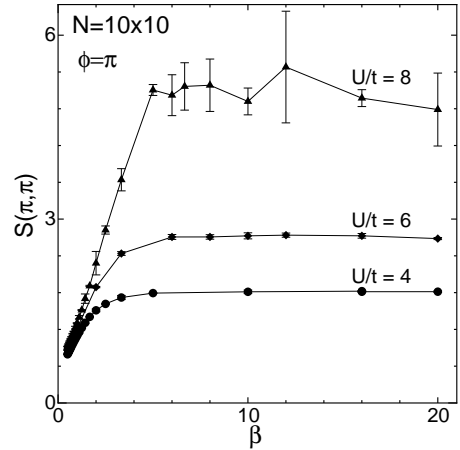,width=6.0cm}
  \caption{
  The antiferromagnetic structure factor $S(\pi, \pi)$
  versus the inverse of temperature $\beta$
  for $\phi=\pi$ on a $N=10 \times 10$ lattice
  with interaction strengths
  $U/t$=
  4 (circles),
  6 (diamonds), and
  8 (triangles).
  The antiferromagnetic correlation enhances as $U/t$ increase.}
  \label{fig:spin-03}
 \end{center}
\end{figure}

  Fig.\ref{fig:spin-03} shows 
the antiferromagnetic structure factor $S(\pi,\pi)$ 
for a variety of $U/t$.
  The antiferromagnetic correlation enhances as $U/t$ increases.  
  As noted above, when $U/t$ is sufficiently large, 
effects of the flux become irrelevant 
and one can expect that 
the antiferromagnetic long-range order appears.
  Due to the numerical difficulties, 
we can not perform simulations 
for stronger interaction ($U/t > 10$) regime.
However, if the interaction is sufficiently strong ($U{\gg}t$),
the model (\ref{eq:model_Hamiltonian}) is essentially 
described by the antiferromagnetic Heisenberg model.
Therefore the antiferromagnetic long-range order
also may appear at some finite value of the interaction strength.

\section{discussion and summary}

  We have studied effects of the interaction in the flux phase.
  The Mott transition is focused using the quantum Monte Carlo method.
  Our results on the charge compressibility shows that
effects of the flux is relevant for small $U/t$, 
while it becomes irrelevant when $U/t$ is sufficiently large.  
The antiferromagnetism,
which is characteristic of the Mott insulator, 
is also strongly suppressed in the weak coupling region.
  This is due to the structure of the low energy excitations 
in the flux phase.
  It implies that 
the flux state with interaction leads to 
a new singular phase for $U < U_{c}$.
  This is in contrast to
the standard two-dimensional Hubbard model.
Effects of the doping is also an interesting future issue
in connection with  the competition with the superconductivity.

\section*{ACKNOWLEDGMENTS}

We are grateful to 
Y.~Morita, 
M.~Yamanaka 
and Y.~Kato 
for helpful comments.  
Y.H is supported in part by 
Grant-in-Aid from the Ministry of Education,
Science and Culture of Japan.  
The computation in this work has been done 
using the facilities of 
the Supercomputer Center, ISSP, University of Tokyo.

\newpage

\end{document}